\documentclass[12pt]{article}
\usepackage{amsmath,amssymb}
\usepackage{hyperref}
\usepackage{graphicx}

\numberwithin{equation}{section} 

\newcommand{\Tr}{{\rm{Tr}}}
\newcommand{\del}{\partial}

\newcommand{\SD}{S_{D3}}
\newcommand{\X}{Y}

\begin{document}


\thispagestyle{empty}

 \renewcommand{\thefootnote}{\fnsymbol{footnote}}
\begin{flushright}
 \begin{tabular}{l}
 {\tt arXiv:0908.3938[hep-th]}\\
 {SNUTP 09-011}
 \end{tabular}
\end{flushright}

 \vfill
 \begin{center}
 {\bfseries \Large Operator with large spin and spinning D3-brane}
\vskip 1.9 truecm

\noindent{{\large
  Dongmin Gang \footnote{arima275(at)snu.ac.kr},
Jae-Sung Park \footnote{parkjs22(at)snu.ac.kr}
and Satoshi Yamaguchi \footnote{yamaguch(at)phya.snu.ac.kr} }}
\bigskip
 \vskip .9 truecm
\centerline{\it Department of Physics and Astronomy,
Seoul National University,
Seoul 151-747, KOREA}
\vskip .4 truecm
\end{center}
 \vfill
\vskip 0.5 truecm

\begin{abstract}
We consider the conformal dimension of an operator with large spin, using a spinning D3-brane with electric flux in $AdS_5 \times S^5$ instead of spinning fundamental string.  This spinning D3-brane solution seems to correspond to an operator made by taking trace in a large symmetric representation.
The conformal dimension, the spin and the R-charge show a scaling relation in a certain region of parameters. In the small string charge limit, the result is consistent with the fundamental string picture. There is a phase transition when the fundamental string charge become larger than a certain critical value; there is no stable D3-brane solution above the critical value.
\end{abstract}

\vfill
\vskip 0.5 truecm

\setcounter{footnote}{0}
\renewcommand{\thefootnote}{\arabic{footnote}}

\newpage

\tableofcontents

\section{Introduction and summary}
It is well known that in four-dimensional gauge theories the
anomalous dimensions of composite operators with large Lorentz spin
$S$ scales logarithmically with the
spin\cite{Callan:1973pu,Belitsky:2003ys,Korchemsky:1995be,Belitsky:2006en}.
In \cite{Gubser:2002tv}, this logarithmic scaling at strong coupling
in $\mathcal{N}=4$ SYM was shown by using AdS/CFT correspondence
\cite{Maldacena:1997re}. They consider a folded string rotating in
$AdS_5$ which is claimed to be dual to an operator of twist two with
large spin in the gauge theory. The classical energy $E$ of the
string (scaling dimension of the operator) scales at large spin as
\begin{align}
E=S+\frac{\sqrt{\lambda}}{\pi}\log S+\ldots,  \label{cusp anomalous}
\end{align}
where $\lambda$ is the 't Hooft coupling. This analysis is extended
to the operator with large $R$-charge $J$ as well as $S$, in the
form of  $\Tr(D^S Z^J)$ (here $D=D_0+D_1$ is the covariant
derivative and $Z$ is a complex scalar) which corresponds to a folded string rotating in
both $AdS_5$ and $S^5$ in \cite{Frolov:2002av}. In the ``long string
limit,''
\begin{align}
 S\gg J  \textrm{ and }  \frac{\sqrt{\lambda}}{\pi J}\log \frac{S}{J}=\textrm{fixed} \label{long string limit},
\end{align}
the classical energy of the string can be written in a closed form \cite{Belitsky:2006en}.
\begin{align}
E=S+J\sqrt{1+ \frac{\lambda}{\pi^2 J^2}\log^2 \frac{S}{J}}+... \label{long string spectrum}
\end{align}
These operators belong to the $SL(2)$ sector in the gauge theory.
Spectrum of such ``long'' operators can be analyzed using integrability of planar $\mathcal{N}=4$ SYM \cite{Beisert:2006ez}. Partial list of the recent development includes \cite{Ryang:2004tq,Kruczenski:2004wg,Eden:2006rx,Sakai:2006bp,Benna:2006nd,Frolov:2006qe,Basso:2006nk,Casteill:2007ct,Alday:2007mf,Kruczenski:2007cy,Basso:2007wd,Freyhult:2007pz,Kostov:2008ax,Kruczenski:2008bs,Bombardelli:2008ah,Fioravanti:2008rv,Basso:2008tx,Fioravanti:2008ak,Buccheri:2008ap,Gromov:2008en,Beccaria:2008nf,Fioravanti:2008bh,Bajnok:2008it,
Ishizeki:2008tx,Fioravanti:2009xn,Freyhult:2009bx,Rej:2009je}.

We want to consider what happens when this macroscopic string is replaced by a D-brane\cite{Callan:1997kz,Gibbons:1997xz}.  There are several studies on this problem with Wilson lines in AdS/CFT \cite{Rey:1998ik,Maldacena:1998im} which originally correspond to macroscopic fundamental strings. When the macroscopic string is replaced by a D-brane with electric flux\cite{Rey:1998ik,Drukker:2005kx,Hartnoll:2006hr,Yamaguchi:2006tq,Gomis:2006sb}, it corresponds to a Wilson line of a higher representation; a D3-brane corresponds to the $k$-th symmetric representation while a D5-brane corresponds to the $k$-th anti-symmetric representation, where $k$ is the string charge of the D-brane with electric flux.

%

Thus, replacing the folded rotating string by a folded rotating D3
or D5-brane, one could analyze the spectrum of the local operators
in the $k$-th symmetric or anti-symmetric representation
\footnote{More precisely $\Tr$ in the fundamental representation is
replaced by the character (or Schur polynomial) of the symmetric or
the anti-symmetric representation.}.  The spectrum of twist two
operators in the $k$-th anti-symmetric representation is studied by
Armoni \cite{Armoni:2006ux} using rotating D5-brane. Another related
study is ``fat magnon''\cite{Hirano:2006ti} which is a D3-brane
version of the macroscopic string solution called ``giant
magnon''\cite{Hofman:2006xt}. Other related solutions are found in
\cite{Hartnoll:2006ib,Drukker:2006zk}.


In this paper, we study the spectrum of the operator $\Tr(D^S Z^J)$
in a symmetric representations  using rotating D3-brane probe. We
find classical D3-brane solutions rotating in both $AdS_5$ and
$S^5$. To use the holographic dictionary for Wilson lines and the
more symmetries, we study the D3-brane counterpart of the ``long
string'' \eqref{long string limit}. In the ``long string'' case, the
folded string touches the boundary of $AdS_5$ (so it represents
Wilson lines) and one more symmetry is enhanced (translation in
$\chi$. See the section \ref{sec-setup}).  As a result, we find
following scaling behavior in certain parameter regime.
\begin{align}
&(E-S)^2 -J^2 = T_3^2 f(\beta,\mu)\log^2 \frac{S}{J}, \label{spectrumofD3}
\\
&\textrm{where, }  \beta= \frac{ J}{T_3 \log \frac{S}{J}}, \quad \mu = 2\pi \frac{\sqrt{\lambda}k}{N}, \quad T_3 = \frac{N}{2\pi^2}.
\end{align}
Which is valid when
\begin{align}
\beta,\mu \textrm{ fixed}, \quad S \gg J,\quad  N\rightarrow \infty, \quad \lambda \rightarrow \infty.  \label{limit of validity}
\end{align}
In small $\beta$ and $\mu$, the function $f$ can be expanded as polynomial in $\beta^2$ and $\mu^2$.
\begin{align}
f(\beta,\mu)  = \mu^2 + c_{2,0}\beta^4 + c_{1,1} \beta^2 \mu^2 + c_{0,2}\mu^4+ \textrm{higher order terms}.
\end{align}
Thus from  \eqref{spectrumofD3}, the anomalous dimension $\gamma: = E-S-J$ can be written as
\begin{align}
\gamma &= J \sum_{m=0}^{\infty} \beta^{2m} \gamma_m (x^2),
\\
&=J\big{[}(\sqrt{1+ x^2}-1)+ \beta^2 ( \frac{c_{2,0}+c_{1,1}x^2+ c_{0,2}x^4}{2\sqrt{1+x^2}}) + o(\beta^4) \big{]}, 
\\
&\textrm{where, } x:=\frac{\mu}{\beta} = \frac{k\sqrt{\lambda}\log(S/J)}{\pi J}.
\end{align}
Note that the double expansion in $\beta^2$ and $x^2$ has same structure with the double expansion in $\frac{1}{N^2}$ and $\lambda$ in the gauge theory side. At planar order (zeroth in $\beta^2$), the anomalous dimension coincide with that of $k$ noninteracting folded strings (compare with \eqref{long string spectrum}).

In some region in  $(\beta, \mu)$, there is no classical D3-brane solution. There exists a critical value for $\mu$ for each $\beta$ above which the classical D3-brane solution does not exist (see Figure \ref{fig1}). This seems to be a similar phenomenon as the phase transition in a symmetric Wilson loop observed in \cite{Hartnoll:2006hr,Okuyama:2006jc,Grignani:2009ua}. It would be interesting further work to check this phase transition in the gauge theory side.

In order to calculate this anomalous dimension in the gauge theory side, one should consider the limit $N\to \infty$ while keeping $\beta,\mu$ finite and $\lambda$ small finite instead of the limit \eqref{limit of validity}. In this limit certain kinds of non-planer diagrams also contribute to the result since $\mu$ kept finite. It will be an interesting future work to consider what kinds of diagrams contribute and what do not.

\section{Set up}
\label{sec-setup}
\subsection{Symmetry, ansatz and boundary condition}
First we will consider the symmetries of ``infinity strings'' in
\cite{Gubser:2002tv} which are dual to twist two operators with
large spin, $S \gg \sqrt{\lambda}$. From those symmetries, we will
find the appropriate ansatz and boundary conditions for the D3-brane
which wraps 4 dimensional submanifold in $AdS_5$ and ends on the two
light-like segments in the $AdS_5$ boundary. Then we will generalize
the ansatz by turning on the angular momentum along $S^5$.

The infinite string solution is given by \cite{Kruczenski:2007cy} \footnote{Actually the folded string world-sheet covers \eqref{infinite string} twice. Thus quantum numbers of the folded string should be doubled if one calculates them using \eqref{infinite string}. This two-foldedness should be taken into account in calculating quantum numbers for folded D3-brane. }
\begin{align}
X_{-1}X_2 -X_0 X_1 =0, \quad X_3=X_4=0 . \label{infinite string}
\end{align}
Here $\{X_\mu\}$ are the Cartesian coordinates of $\mathbb{R}^{2,4}$ where the $AdS_5$ is embedded.
In the global coordinates $\{\tilde{\tau},\tilde{\rho},\Omega_i\}$, $(i=1,2,3,4)$ for $AdS_5$,
\begin{align}
X_{-1}=\cosh \tilde{\rho} \cos \tilde{\tau}, \quad X_{0}=\cosh \tilde{\rho} \sin \tilde{\tau}, \quad X_i =\sinh \tilde{\rho} \Omega_i, \quad \sum_{i=1}^{4}\Omega_i^2 =1,
\end{align}
the boundary is located at $\tilde{\rho}\rightarrow \infty$. The infinite string  ends on  the following two light-like lines at the boundary.
\begin{align}
\tilde{\tau} = \tilde{\varphi}\quad  \textrm{or} \quad \tilde{\tau}=\tilde{\varphi}+\pi , \quad \Omega_3^2+\Omega_4^2=0. \quad \textrm{where } \tilde{\varphi} = \arctan\frac{\Omega_2}{\Omega_1} . \label{light-like segments}
\end{align}
These two light-like Wilson lines preserve three symmetries of
$SO(2,4)$. These symmetries are more manifest in the $AdS_3 \times
S^1$ foliation of $AdS_5$.
\begin{align}
&(X_{-1},X_0,X_1,X_2)=\cosh \zeta (x_{-1},x_0,x_1,x_2), \quad -x_{-1}^2-x_0^2+x_1^2+x_2^2=-1,\nonumber
\\
&(X_3,X_4)=\sinh \zeta (x_3,x_4), \quad x_3^2+x_4^2=1,\nonumber
\\
&ds^2 (AdS_5) = \cosh^2 \zeta ds^2 (AdS_3)+\sinh^2 \zeta d\psi^2+d\zeta^2. \label{foliation}
\end{align}
We will use two coordinate systems for $AdS_3$, $\{u,\chi,\sigma\}$ and $\{\tau,\rho,\varphi\}$. See Appendix \ref{app-coordinates}. The infinite string \eqref{infinite string} stretches along $u,\chi$ directions and located at $\zeta=0,\sigma=0$.
And the three symmetries correspond to translations in $u,\chi$ and $\psi$ \cite{Alday:2007mf}. Besides these continuous symmetries, there is an additional $\mathbb{Z}_2$
symmetry, $\sigma \leftrightarrow -\sigma$.

We will consider the D3-brane motion described by the DBI+WZ action
\begin{equation}
\begin{split}
 &\SD=T_3\int d^4 y L=T_3\int d^4 y(L_{DBI}+L_{WZ}),\\
& L_{DBI}=-\sqrt{-\det H},\qquad H_{\alpha\beta}:=
G_{MN}(\X)\frac{\del \X^M}{\del y^{\alpha}}\frac{\del \X^N}{\del y^{\beta}}
 +F_{\alpha\beta},\\
& L_{WZ}=-a C_{M_1\dots M_4}\frac{\del \X^{M_1}}{\del y^{\alpha_1}}\dots
\frac{\del \X^{M_4}}{\del y^{\alpha_4}}\epsilon^{\alpha_1 \dots \alpha_4}\frac{1}{4!},
\end{split}\label{DBIWZ}
\end{equation}
where $\X^M,\ (M=0,\dots,9)$ denote the space-time coordinates and $y^{\alpha}, \ (\alpha=0,1,2,3)$ are the D3-brane world-volume coordinates. $F_{\alpha\beta}$ is the world-volume gauge flux. $a$ is $\pm 1$ depending on the choice of the orientation. The D3-brane tension $T_3$ is related to $N$ by $T_3=\frac{N}{2\pi^2}$ in our unit (AdS radius)$=1$.

We are going to find classical D3-brane solution which preserves the three symmetries and ends on the light-like segments \eqref{light-like segments} at the $\mathbb{R}\times S^3$ boundary.
{}From three continuous symmetries, the ansatz for D3-brane is
($\{u,\chi,\psi,y \}$ are the world-volume coordinates)
\begin{align}
F = b\, du\, d\chi, \quad \sigma = \sigma(y), \quad \zeta = \zeta(y). \label{ansatz 1}
\end{align}
To preserve the $\mathbb{Z}_2$ symmetry ($\sigma \leftrightarrow -\sigma$), we impose the following.
\begin{align}
\frac{d\zeta}{d\sigma} =0 , \quad \textrm{when } \sigma=0. \label{ansatz 2}
\end{align}
When $\zeta=0$ where the size of $S^1$ shrinks, there may be conical
singularity. To avoid this, we impose the following condition.
\begin{align}
\frac{d\sigma}{d\zeta} =0 , \quad \textrm{when } \label{ansatz 3}
\zeta=0.
\end{align}
In the $AdS_3 \times S^1$ foliation, the $\mathbb{R} \times S^3$ boundary of $AdS_5$ is located at $\cosh^2 \zeta \cosh^2 \rho \rightarrow \infty$, or equivalently \begin{align}
\rho\rightarrow \infty \quad \textrm{or} \quad \zeta\rightarrow \infty. \label{boundary}
\end{align}
And the two light-like lines at the boundary \eqref{light-like segments} becomes
\begin{align}
&\tau = \varphi \quad \textrm{or} \quad \tau= \varphi +\pi, \nonumber
\\
&\frac{X_3^2+X_4^2}{X_{-1}^2+X_0^2} = \frac{\sinh^2 \zeta}{\cosh^2\zeta \cosh^2 \rho} \rightarrow 0 . \label{light-like segments 2}
\end{align}
Under the ansatz \eqref{ansatz 1}, the D3-brane ends on the two segments \eqref{light-like segments 2} at the boundary \eqref{boundary} if and only if
\begin{align}
\sigma(y), \zeta(y) = \textrm{finite}. \label{ansatz 4}
\end{align}
Equations \eqref{ansatz 1},\eqref{ansatz 2},\eqref{ansatz 3},\eqref{ansatz 4} are the summary of ansatz and conditions for D3-brane rotating in $AdS_5$. These can be generalized by turning on the angular momentum along $S^5$:
\begin{align}
\theta = \nu u.
\end{align}
Here $\theta$ is the coordinate of a great circle of $S^5$. Under these ansatz, the D3-brane action \eqref{DBIWZ} becomes
\begin{align}
 \SD=T_3\int du d\chi d\psi dy L,\qquad \nonumber
L=L_{DBI}+L_{WZ},\\
L_{DBI}=-\sqrt{\sinh^2\zeta(\cosh^4 \zeta\cosh^2 2\sigma-b^2-\nu^2\cosh^2\zeta)(\cosh^2\zeta \sigma'^2+\zeta'^2)}, \nonumber\\
L_{WZ}=-a (\cosh^4 \zeta-1)\cosh 2\sigma \sigma' . \label{DBI action}
\end{align}
Here $a$ is $\pm 1$ depending on the choice of the orientation. Under these ansatz, the equation of motion for the world-volume gauge field is automatically satisfied.

\subsection{Quantum numbers}
Here in this subsection we will obtain the expression of the conserved charges: the energy $E$, the spin $S$, the R-charge $J$, and the string charge $k$. First three charges $E,S,J$ are calculated as the Noether charges from the spacetime isometry. Later we will calculate $k$ by taking variation by NSNS B-field.

For a Killing vector $\xi^{M}$ in $AdS_5\times S^5$ and a small parameter $\epsilon$, there is a symmetry of the action \eqref{DBIWZ}.  Since this isometry also preserves the RR5-form field strength $F_5$, the variation of 4-form potential should be written as
\begin{align}
 \delta C_{4}=\epsilon d\Lambda_3,
\end{align}
where $\Lambda_3$ is a 3-form. The variation of the Lagrangian becomes
\begin{align}
 \delta L=\epsilon \del_{\alpha}\left[-a \frac1{3!}\epsilon^{\alpha\alpha_2\alpha_3\alpha_4}\del_{\alpha_2}\X^{M_2}\del_{\alpha_3}\X^{M_3}\del_{\alpha_4}\X^{M_4}\Lambda_{M_2M_3M_4}\right]
 =:\epsilon\del_{\alpha}R^{\alpha}.
\end{align}
The Noether current $j^{\alpha}$ and the Noether charge $Q$ for this symmetry is written as
\begin{align}
& j^{\alpha}=\frac{\del L}{\del(\del_{\alpha}\X^{M})}\xi^{M}-R^{\alpha} \label{current},\\
& Q=T_3\int d^3 y\; j^{0} \label{charge}.
\end{align}

We only need to consider DBI-term in the action because we are considering folded D3-brane solution. Actually the terms in eq.\ \eqref{current} which come from the WZ-term cancel since two D3-branes have the opposite sign of the WZ-term to each other. The derivative of the DBI-term is given by
\begin{align}
 \frac{\del L_{DBI}}{\del(\del_{\alpha}\X^{M})}=
-\sqrt{-\det H}(H^{-1}_{sym})^{\alpha\beta}G_{MN}\del_{\beta}\X^{N},
\end{align}
where $H^{-1}_{sym}$ is the symmetric part of the inverse matrix of $H$.

We take $u$ as the world-volume time.
For the R-charge $J$ the Killing vector is $\xi_{J}=\del/\del \theta$. The Noether charge is given as
\begin{align}
 &J=T_3\int d\chi \int d\psi \int dy\, j_{J}^{u}=2\chi_0 T_3 \beta,\\
 &\beta:=\int dy\frac{4\pi\nu (\cosh^2\zeta \sigma'^2+\zeta'^2)\sinh\zeta\cosh^2\zeta}{\sqrt{(\cosh^4\zeta\cosh^22\sigma-b^2-\nu^2\cosh^2\zeta)(\cosh^2\zeta \sigma'^2+\zeta'^2)}},\label{beta}
\end{align}
where $\chi_0$ is the cut-off of the $\chi$ integral; $\chi$ is limited to $-\chi_0 \le \chi \le \chi_0$. As the same way, the Killing vector for $E-S$ is
$\xi_{E-S}=-\del/\del{\tau}-\del/\del{\varphi}=-\del/\del u$ (see eqs.\ \eqref{Killing-E} and \eqref{Killing-S}), and the Noether charge is obtained as
\begin{align}
& E-S=2\chi_0 T_3 \alpha,\\
&\alpha:=\int dy\frac{4\pi(\cosh^2\zeta \sigma'^2+\zeta'^2)\sinh\zeta\cosh^4\zeta\cosh^2 2\sigma}{\sqrt{(\cosh^4\zeta\cosh^22\sigma-b^2-\nu^2\cosh^2\zeta)(\cosh^2\zeta \sigma'^2+\zeta'^2)}}.\label{alpha}
\end{align}

On the other hand, for the spin $S$, the components of the Killing vector behave as $\xi_{S}\sim e^{2\chi}$ in large $\chi$ (see eq.\ \eqref{Killing-S}). Thus the charge $S$ after integral over $\chi$ behaves as
\begin{align}
 S\sim T_3 e^{2\chi_0}, \text{ or } 2\chi_0\sim \log \frac{S}{J}.
\end{align}
As a result we obtain the scaling behavior
\begin{align}
 (E-S)^2-J^2=T_3^2(\alpha^2-\beta^2)\log^2 \frac{S}{J}. \label{scaling behavior}
\end{align}

Let us turn to the string charge $k$. For a variation of B-field
$\delta B_{u\chi}$, the variation of the action and the string
charge $k$ are related as ($\alpha'=\frac{1}{\sqrt{\lambda}}$ is the
slope parameter in our unit.)
\begin{align}
 \delta \SD=\frac{k}{2\pi \alpha'}\int du d\chi \delta B_{u\chi}.
\end{align}
Hence the string charge $k$ is expressed as
\begin{align}
 k=&2\pi \alpha' T_3 \int d\psi \int dy \frac{\del L}{\del b}
=\frac{N}{2\pi\sqrt{\lambda}}\mu,\\
\mu:=&\int dy\frac{4\pi b \sinh\zeta(\cosh^2\zeta \sigma'^2+\zeta'^2)}{\sqrt{(\cosh^4\zeta\cosh^22\sigma-b^2-\nu^2\cosh^2\zeta)(\cosh^2\zeta \sigma'^2+\zeta'^2)}}.\label{gamma}
\end{align}
The scaling function $f(\beta,\mu)$ in eq.\ \eqref{spectrumofD3} is obtained from \eqref{scaling behavior} by expressing $\alpha^2-\beta^2$ in terms of $\beta$ and $\mu$.
\section{Numerical analysis}

So far we describe the general procedure for obtaining a D3-brane
solution which is dual to the composite operator $\Tr(D^S Z^J)$ in
symmetric representations. In this section, we will find numerical
solutions and analyze its phase structure and energy spectrum.
\subsection{Phase structure}
\begin{figure}
\begin{center}
  \includegraphics[width=6truecm]{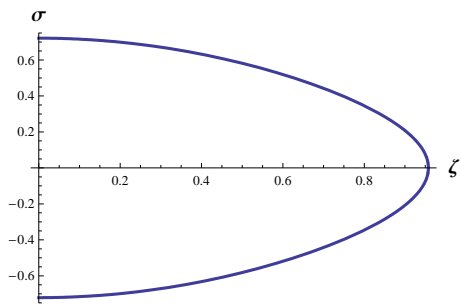}
\end{center}
 \caption{D3-brane example ($\nu=0.999 , b=0.1$).}\label{fig1}
\end{figure}

\begin{figure}
\begin{center}
  \includegraphics[width=6truecm]{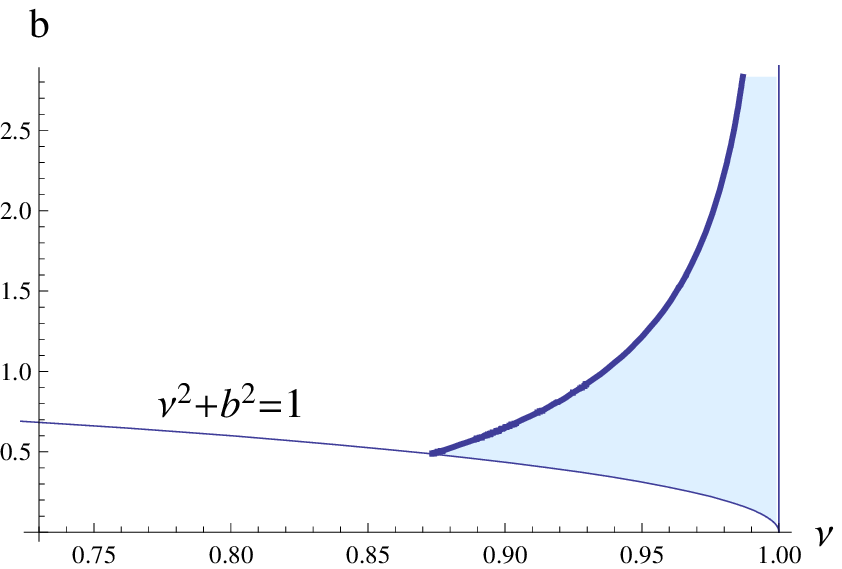}
  \includegraphics[width=6truecm]{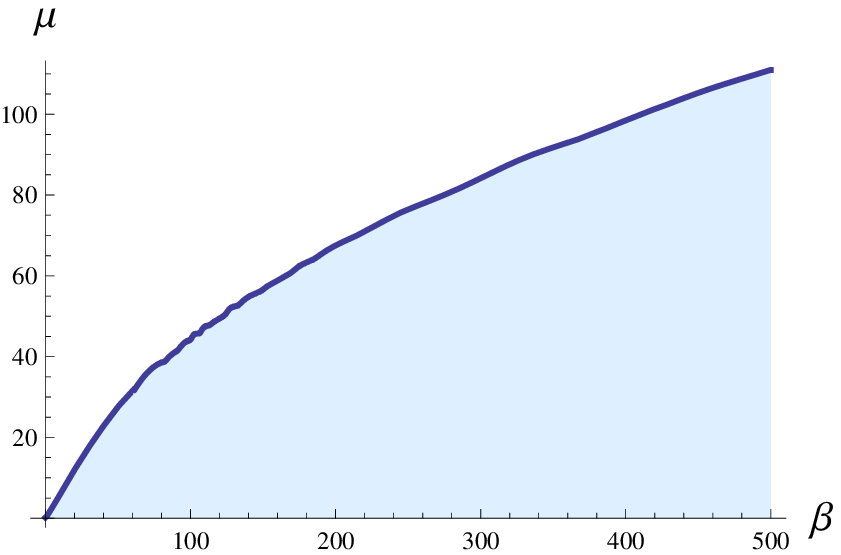}
\end{center}
 \caption{left: the region stable solutions exit, right: ($\beta,\mu$) of the solutions fill the colored region.   }\label{fig2}
\end{figure}

The equation of motion derived from the action \eqref{DBI action} is
too complicated to solve it analytically. Thus, we find solutions
numerically. The solution which satisfies the conditions
\eqref{ansatz 2},\eqref{ansatz 3},\eqref{ansatz 4} looks like an
ellipse in $(\zeta,\sigma)$-plane (see the Figure \ref{fig1}).
For some values of $(\nu,b)$, there are several solutions. But if we
impose stability condition\footnote{We check the stability
numerically. We consider several small fluctuations $\{ \delta
\sigma, \delta \zeta \}$ around a solution. If the solution maximize
the Lagrangian $\int d\chi d\psi d y L$ under the fluctuations, then
it is considered as a stable one. }, only one or no solution survives.
And for some other values of $(\nu , b)$, there's no solution (even
unstable one). Figure \ref{fig2} shows the region in $(\nu,b)$ where the stable solutions
exist.
The region is surrounded by following three curves.

\begin{itemize}

\item $\nu^{2}+b^{2}=1$.
\\
To avoid the Lagrangian \eqref{DBI action} being an imaginary number, there's a lower bound for the size of solutions.
\begin{align}
r := \sqrt{\zeta^2+\sigma^2} \geq \frac{1}{2}\mathrm{arccosh}(\sqrt{\nu^{2}+b^{2}})
\end{align}
When $\nu^{2}+b^{2}$ approaches to 1, the bound becomes smaller and stable solutions tend to shrink to the point $r=0$. Accordingly, the physical quantities $(\beta,\mu)$ of the solution become smaller.

\item $\nu = 1$.

If $\nu \geq 1$, there's no solution except unstable one.  Stable solutions (and its $\beta,\mu$) in the colored region become infinity when $\nu \rightarrow 1$. This bound for the angular velocity in $S^5$ direction also exists for the folded string solution case \cite{Frolov:2002av}.\footnote{ We fix $\frac{\omega}{\kappa} =1 $ and $\frac{\nu}{\kappa}$ in \cite{Frolov:2002av} corresponds to $\nu$ in this paper.}

\item The upper curve.

We cannot find analytic expression for this curve.  Just below the curve there are two solutions (1 stable + 1 unstable). The two solutions get closer to each other when approaching the upper curve and disappear simultaneously above the curve. This curve is mapped to the upper curve in the $(\beta,\mu)$ plane via stable solutions.  It suggests that there's some phase transition across the curve. This result requires further study to understand the phase transition in the gauge theory side.
\end{itemize}

\subsection{Expansion in small $\beta, \mu$.}
The energy spectrum \eqref{spectrumofD3}, which is valid in the limit \eqref{limit of validity}, is wholly determined if we find expression $f:=\alpha^2-\beta^2$ in terms of $\beta,\mu$. Although we cannot find its full analytic expression, we suggest the form of series expansion and obtain the exact values of the coefficients at the lowest order. Higher order coefficients can be obtained numerically.

Consider the limit $(\nu,b)$ approaching to the curve $\nu^2+b^2=1$. In the limit, as mentioned above, the stable solution (and its $\beta,\mu$) become smaller. From the expression for $\alpha,\beta,\mu$ in section 2 and the fact that $\zeta,\sigma$ is very small, one can see that
\begin{align}
\frac{f}{\mu^2}=\frac{\alpha^2 - \beta^2}{\mu^2} \simeq \frac{1-\nu^2}{b^2} \rightarrow 1 \label{alpha,beta->0 limit case}
\end{align}
in the limit. This result gives
\begin{align}
(E-S)^2 -J^2 = k^2 \frac{\lambda}{\pi^2}\log^2 (\frac{S}{J}) \quad \textrm{when } \beta, \mu \rightarrow 0 .
\end{align}
This is nothing but the spectrum of $k$ noninteracting  folded strings!(cf. \eqref{long string spectrum}).

Assuming the $f(\beta,\mu)$ is analytic near the origin $(\beta,\mu)=(0,0)$, we propose following expansion
\begin{align}
f(\beta,\mu) = \sum_{m,n} c_{m,n}\beta^{2m} \mu^{2n}, \quad m,n \geq 0.  \label{expansion of f}
\end{align}
Here we use the fact that $f(\beta,\mu)$ is even function in both $\beta$ and $\mu$.\footnote{When $\nu \leftrightarrow -\nu$, e.o.m does not change and the stable solution remains same. So were $\alpha,\mu$. But $\beta$  changes its sign \eqref{beta}. Similar argument hold for the $b \leftrightarrow -b$ case ( in this case, $(\alpha,\beta)$ remains same but $\mu$ changes its sign. ).} And eq.\ \eqref{alpha,beta->0 limit case} imply that
\begin{align}
c_{0,0}=0,\quad  c_{1,0}=0, \quad c_{0,1}=1.
\end{align}
Numerically, we check the expansion \eqref{expansion of f} up to fourth power of $\beta,\mu$ and obtain the value of $c_{2,0},c_{1,1},c_{0,2}$.
\begin{figure}
\begin{center}
  \includegraphics[width=7truecm]{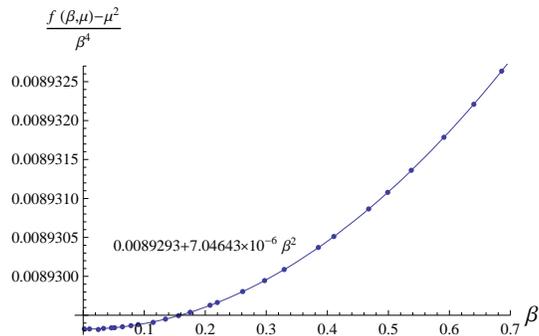}
\end{center}
 \caption{Along $\mu=0.306857 \beta$ we plot the graph. Its behavior agree with what we expected from \eqref{expansion of f}.}\label{fig3}
\end{figure}
\begin{align}
c_{2,0}=0.0084\ldots, \quad c_{1,1}=0.0074\ldots, \quad c_{0,2}=-0.023\ldots .
\end{align}
For small $(\beta,\mu)$, $f(\beta,\mu)$ is well approximated by this
expansion as shown in Figure \ref{fig3}.
\subsection*{Acknowledgments}
We would like to thank Luis F. Alday and Soo-Jong Rey for useful discussions and comments. The discussions during ``APCTP Focus Program on Liouville, Integrability and Branes (5)'' May 17-30, 2009 at APCTP and ``Summer Institute 2009'' Aug. 3-13, 2009 at Yamanashi were useful to finish this work.
This work was supported in part by KOFST BP Korea Program, KRF-2005-084-C00003, EU FP6 Marie Curie Research and Training Networks MRTN-CT-2004-512194 and HPRN-CT-2006-035863 through MOST/KICOS.

\appendix
\section{Coordinates of AdS$_3$}
\label{app-coordinates}
We mainly use the coordinates of the $AdS_3\times S^1$ foliation of $AdS_5$ \eqref{foliation}.
The coordinates of $AdS_3$ appeared in \cite{Alday:2007mf} is convenient for our purpose. We summarize its relation to the usual global coordinates.
The coordinates $(u,\chi,\sigma)$ are given by
\begin{equation}
\begin{aligned}
 x_{-1}&=\cos u \cosh \sigma \cosh \chi-\sin u \sinh \sigma \sinh \chi,\\
 x_{0}&=\sin u \cosh \sigma \cosh \chi+\cos u \sinh \sigma \sinh \chi,\\
 x_{1}&=\cos u \cosh \sigma \sinh \chi-\sin u \sinh \sigma \cosh \chi,\\
 x_{2}&=\cos u \sinh \sigma \cosh \chi+\sin u \cosh \sigma \sinh \chi.
\end{aligned}
\end{equation}
The metric in this coordinates is written as
\begin{align}
 ds^2(AdS_3)=-du^2+d\chi^2-2 \sinh 2\sigma\, du d\chi+d\sigma^2.
\end{align}

On the other hand, the global coordinates $(\tau,\rho,\varphi)$ parametrize the
$AdS_3$ as
\begin{equation}
\begin{aligned}
 x_{-1}=\cosh \rho \cos \tau,\qquad x_0=\cosh \rho \sin \tau,\\
 x_{1}=\sinh \rho \cos \varphi,\qquad x_2=\sinh \rho \sin \varphi.
\end{aligned}
\end{equation}
The metric in this global coordinates is written as
\begin{align}
 ds^2(AdS_3)=-\cosh^2 \rho d\tau^2 +d\rho^2+\sinh^2\rho d\varphi^2.
\end{align}

These two coordinate systems are related as
\begin{align}
 &\sinh\rho=\sqrt{\cosh^2\sigma \sinh^2 \chi+\sinh^2\sigma \cosh^2\chi},\\
 &\tan \tau=\frac{\tan u+\tanh \sigma\tanh \chi}{1-\tan u \tanh \sigma \tanh \chi},\\
 &\tan \varphi=\frac{\tanh \sigma+\tan u \tanh \chi}{\tanh\chi-\tan u\tanh\sigma},
\end{align}
or
\begin{align}
  &\sinh2\sigma=\sinh 2\rho \sin(\varphi-\tau),\\
  &\sinh2\chi=\frac{\sinh 2\rho \cos(\varphi-\tau)}{\sqrt{1+\sinh^2 2\rho \sin^2(\varphi-\tau)}},\\
 &e^{4iu}=e^{2i(\tau+\varphi)}\frac{\cos(\varphi-\tau)-i\cosh2\rho\sin(\varphi-\tau)}{\cos(\varphi-\tau)+i\cosh2\rho\sin(\varphi-\tau)}.
\end{align}

The Killing vectors corresponding to the energy and the angular momentum are given in the $(u,\chi,\sigma)$ coordinates as
\begin{align}
 \del_{\tau}
=&\frac12\left(1+\frac{\cosh2\chi}{\cosh 2\sigma}\right)\del_u
+\frac12\cosh2\chi \tanh 2\sigma \del_{\chi}
-\frac12\sinh 2\chi \del_{\sigma},\label{Killing-E}\\
 \del_{\varphi}
=&\frac12\left(1-\frac{\cosh2\chi}{\cosh 2\sigma}\right)\del_u
-\frac12\cosh2\chi \tanh 2\sigma \del_{\chi}
+\frac12\sinh 2\chi \del_{\sigma}.\label{Killing-S}
\end{align}

\providecommand{\href}[2]{#2}\begingroup\raggedright\endgroup

\end{document}